\documentclass{ws-ijmpa}
\usepackage{psfrag}

\begin{document}

\markboth{Chao-Hsi Chang (Zhao-Xi Zhang)} {Production and decay of
the meson $B_c$}

\catchline{}{}{}{}{}
%
%%%%%%%%%%%%%%%%%%%%%%%%%%%%%%%%%%%%%%%%%%%%%%%%%%%%%%%%%%%%%%%%%%%%

\title{Production and decay of the meson $B_c$}

\author{\footnotesize Chao-Hsi Chang (Zhao-Xi Zhang)\footnote{In
collaboration with Y.-Q. Chen, X.-Q. Li, C.-D. Lu, C.-F. Qiao,
G.-L. Wang, J.-X. Wang, X.-G. Wu, et al.}}

\address{ CCAST (World Laboratory), P.O. Box 8730, Beijing 100080,
China\\ Institute of Theoretical Physics, Chinese Academy of
Sciences, Beijing 100080, China}

\maketitle

%\pub{Received (Day Month Year)}{Revised (Day Month Year)}

\begin{abstract}
Studies on the decay and production of $B_c (B^*_c)$ meson are
briefly reviewed. Considering RUN-II of Tevatron and the schedule
of LHC, the theoretical studies of $B_c$ meson will jump to a new
stage not only for itself but also to implement the studies of the
heavy quarkonia etc.

\keywords{$B_c$ meson; Decay; Production.}
\end{abstract}

\section{Introduction (the meson $B_c$)}
The meson $B_c$ is the ground state of the heavy-flavored binding
system $(c\bar{b})$, and it is the unique `double heavy-flavored'
meson in Standard Model (SM) and is stable for strong and
electromagnetic interactions. The binding interaction for the
system is similar to the case of heavy quarkonium $(c\bar{c})$ and
$(b\bar{b})$. Its mass and the spectrum of the binding system can
be computed by potential model\cite{potential,prod},
PNRQCD\cite{prod,pnrqcd} and lattice QCD\cite{latt} etc. The
results are in the region, $m_{B_c}\simeq 6.2\sim 6.4$GeV. Its
lifetime was estimated in terms of the effective theory of weak
interaction and by applying the effective Lagrangian to the
inclusive processes of $B_c$ decays\cite{prod,dec,bigi,changli}.
According to the estimates, the lifetime $\tau_{B_c}\simeq 0.4$ps,
a typical one for weak interaction via virtual $W$ boson. Hence
with such a long lifetime, the vertex detection is very useful in
its experimental observations. $B_c$ meson has been observed by
CDF\cite{CDF,CDF1} and D0\cite{D0} already, and so far the
observations are consistent with theoretical predictions within
theoretical uncertainties and experimental errors. \vspace{-4mm}

\section{Decay}
Since the system $(c\bar{b})$ carries two heavy flavors
explicitly, its excited states decay to the ground state $B_c$ by
electromagnetic and/or strong interaction directly or in a cascade
way according to the available phase space with almost 100\%
possibility, while the ground state $B_c$ decays via weak
interaction only. Here I shall concentrate on the weak decay of
the ground state meson $B_c$.

There are two `spectator ways' in the weak decay of $B_c$ meson:
$c$-quark decays with $\bar{b}$-quark as a `spectator' and
$\bar{b}$-quark decays with $c$-quark as a `spectator'. The
interesting aspect of them is that the rates of these two `ways'
are competitive in magnitude: the CKM matrix elements have
$|V_{cb}|\ll |V_{cs}|$, that is in favor of the $c$-quark decay
greatly, whereas the phase space factor is proportional to the
fifth power of the mass of the decay fermion for a weak decay via
a virtual $W$-boson to a massless three-body final state, and
$m_b^5\gg m_c^5$ for the two flavor, that compensates the CKM
matrix element factor a lot. In fact, besides these two, there is
one more important decay way only for the $B_c$ meson decay i.e.
the annihilation of $c$ and $\bar{b}$. Here, when we talk about
the three `ways' individually, it means that we have ignored the
interferences among the amplitudes for the three `ways' in a
moment. The interferences among them actually are not very
great\cite{changli}, therefore, the aspect obtained by the
individual consideration is kept, so $B_c$ has very rich weak
decay channels with comparable decay branching
ratio\cite{prod,dec,bigi,changli}. Furthermore, by measuring the
decay product, one may precisely know the specific decay is caused
by $c$-quark decay or by $\bar{b}$-quark decay. Hence we may study
the two heavy flavors $c$ and $b$ simultaneously just with one
meson $B_c$, and may gain some advantage in comparative studies of
the two flavors e.g. in estimating and measuring the ratio of CKM
matrix elements $\frac{V_{cb}}{V_{cs}}$, because there is a
cancellation for theoretical uncertainties and experimental
systematic errors.

Of the $B_c$ decay, the rate of the decay caused by $c$-quark, in
fact, is bigger than that caused by $\bar{b}$-quark. As a
consequence, even though the available phase space for $c$-quark
decay is comparatively small, the decay $B_c\to B_s(B_s^*)+\cdots$
still has quite great branching ratio: $Br_{B_c\to B_s+\cdots}\geq
50\%$\footnote{Note that here the indirect cascade decays, such as
$B_c\to B_s^*+\cdots$ and $B_s^*\to B_s+\cdots$, are taken into
account too.}. Therefore, `$B_c$ decay to $B_s$' can be used as a
$B_s$ generator (source) potentially if the $B_c$ is produced
numerously. Furthermore, the $B_s$ meson generated from $B_c$
decay is `tagged' explicitly by $B_c$ charge, since $B_c$ carries
positive charge.\vspace{-5mm}

\subsection{Lifetime and Inclusive decays}

To estimate the lifetime of $B_c$, the most reliable way is to
compute its inclusive decays with optics theorem and effective
Lagrangian for weak interaction. It is because that in this way
the non-spectator effects can be taken into account and the
non-perturbative effects are factorized out clearly.

The two spectator components for $B_c$ decay, which are similar to
that of $B$ and $D$ decay accordingly, contribute to the lifetime
with the partial rates: $$\Gamma(b\rightarrow
c)=\sum_{l=e,\mu,\tau}\Gamma^{sl}_{b\rightarrow cl\nu}+
\sum_{q=u,d,s,c}\Gamma^{nonl}_{b\rightarrow c\bar qq}$$ i.e.
$\bar{b}$-quark decay with $c$-quark as spectator, and
$$\Gamma(c\rightarrow s)=\sum_{l=e,\mu}\Gamma^{sl}_{c\rightarrow
s\bar l\nu}+ \sum_{q=u,d,s}\Gamma^{nonl}_{c\rightarrow s\bar qq}$$
i.e. $c$-quark decay with $\bar{b}$-quark as spectator. Each of
them includes both of the semileptonic and nonleptonic decays.

In $B_c$ decay, the annihilation component is another important
one and its contributions are different from those in $B$ and $D$
decays. The so-called weak annihilation (via $W$ boson) component
contains $\Gamma^{WA}_{tree}$, $\Gamma^{WA}_{penguin}$ and
$\Gamma^{WA}(B_c\rightarrow\tau\nu_{\tau})$ which correspond to
the non-leptonic decay induced by the `tree' and `penguin' parts,
and the pure leptonic (PL) decay\footnote{Since the helicity
suppression, here the pure leptonic decays $B_c\to l(e,\mu)+\nu$
are ignored.} respectively. The so-called Pauli interference
components (interferences among the spectator components and
annihilation component) contain the `tree' part
$\Gamma^{PI}_{tree}$ and the `penguin' part
$\Gamma^{PI}_{penguin}$ respectively. The total width of $B_c$
should be the sum of the partial widths $\Gamma=\Gamma^{c\to
s}+\Gamma^{\bar{b}\to\bar{c}}+\Gamma^{WA}+\Gamma^{PI}$ with
$\Gamma^{PI}=\Gamma^{PI}_{tree}+\Gamma^{PI}_{penguin}$ and
$\Gamma^{WA}=\Gamma^{WA}_{tree}+\Gamma^{WA}_{penguin}+
\Gamma^{WA}(B_c\rightarrow\tau\nu_{\tau})$.

\vspace{-4mm}
\begin{table}[h]
\tbl{The lifetime and inclusive branching ratios for $B_c$ meson}
{\begin{tabular}{c c c c c c c c}\hline\hline
%%%%%%    1 row
$f_{B_c}$         &$\tau_{B_c}$                   & $Br(\bar{b})$
& $Br(c)$ & $Br^{WA}$           &$Br^{PI}$
                  & $Br(\tau\nu)$             & $Br^{sl}$    \\ \hline
$.44$         &$0.362$ &$22.8$                  & $70.9$
                 &$13.4$                  & $-7.1$
                 &$2.8$        & $8.7$  \\  \hline
$.50$         &$0.357$ &$22.4$                  &$69.7$
                 &$16.9$                  &$-9.0$
                 &$3.6$         &$8.4$ \\   \hline\hline
\end{tabular}
}
\end{table}
\vspace{-4mm}
\begin{minipage}{119mm}
\noindent {\footnotesize In Table. 1, $\tau_{B_c}$ means the
lifetime in unit $ps$; $f_{B_c}$ is the decay constant in unit
GeV; $Br(\bar{b})$ is the branching ratio of the $\bar{b}$-quark
inclusive decay to $c$-quark and the $c$ inside $B_c$ as the
spectator; $Br(c)$ is the branching ratio of the $c$-quark decay
to $s$-quark and the $\bar{b}$ inside $B_c$ as the spectator;
$Br(\tau\nu)$ is that of the pure leptonic decay $B_c\to \tau\nu$
(without Pauli interference). Owing to the interference,
$\Gamma^{PI}$ is negative, so the value of $Br^{PI}$. All of the
branching ratios in the table are in percentage.}
\end{minipage}\vspace{2mm}

When determining the necessary input parameters, the authors of
Ref.\cite{changli} further considered the measured lifetimes of
$B$ mesons and $D$ mesons as well. As typical ones, here I quote
the values of the theoretical estimate on the $B_c$ lifetime from
Ref.\cite{changli} into Table 1\footnote{In Ref.\cite{changli},
there is a typo that the unit of $\Gamma(\tau\nu)$ should be
$0.1\cdot ps^{-1}$ instead of $ps^{-1}$.}. Due to the
uncertainties in treating the non-perturbative matrix elements in
the estimates, there are some disagreements in the theoretical
estimates on the lifetime which can be found in Ref.\cite{prod},
but at the present stage, all of them agree with the experimental
measurements\cite{CDF,D0} within the experimental errors.
\vspace{-4mm}

\subsection{Pure leptonic decay and its escape from the
chiral-suppression by radiation}

The meson $B_c$ can decay by weak annihilation. Of them, the pure
leptonic decay is of specially interest due to the fact that it
can be used to measure the decay constant $f_{B_c}$ and there is
no strong interaction in final state of the decay and the induced
`penguin' in the effective Lagrangian plays no role to it.

The vector left-hand nature of the weak interaction causes the
so-called chiral suppression. Indeed, due to the suppression, the
decays $B_c\to \bar{l}\nu_l (l=e,\mu)$ are negligible small
because $m_l$ is negligible small. Whereas, if there is additional
particle such as one photon or one gluon in the final state, then
the decay may escape from the chiral suppression. When there is an
additional photon in final state, due to the `escape', the
branching ratio is enhanced by several magnitude order e.g.
$Br(Bc\to e^+\nu_e)+Br(B_c\to e^+\nu_e\gamma) \simeq (10^{4}\sim
10^5) \cdot Br(Bc\to e^+\nu_e)$, and with a hard photon the
absolute branching ratio still can be sizable as $10^{-5}\sim
10^{-4}$. Detail analysis about the suppression and enhancement
can be found in Ref.\cite{changgl}.

Moreover if there is an additional gluon (which fragments to light
hadrons) to the neutrino and charged lepton in the final state,
the decay not only escapes from the chiral suppression but also
can be used as a test of the color-octet components in the meson
$B_c$. It is interesting to point out that such decay can be used
to measure the color-octet components in the meson $B_c$ indeed,
if the color-octet components in the meson are predicted as that
by the `scaling-rule' of NRQCD. When the meson is produced
numerously, as long as more attention is payed to the region of
the decay phase-space that is close to the end point of the
charged lepton (the final state of the decay contains the charged
lepton, light hadrons and missing neutrino only) in the
measurements, the color-octet components can be measured. In the
`end point' region of the decay the contributions from the
color-octet components become greater than those from the
color-singlet components. The quantitative computations on the
decays of $B_c$ to leptons with light hadrons (which are
fragmented from one or two gluons) and the conclusion about the
color-octet component measurements can be found in
Ref.\cite{chwu-oct}.\vspace{-4mm}

\subsection{Semileptonic and Nonleptonic decays}

The essential factor which must be computed in nonleptonic decays
with naive factorization\footnote{Naive factorization can be
applied to nonleptonic decays only for leading-order estimates.}
and in semileptonic decays may be attributed to a matrix element
of the relevant current as $<P'(B_f)|J_\mu|P(A_i)>$ (Fig.1), where
the particle $A_i$ in initial state is $B_c$, and the particle
$B_f$ in final state is one of the particles $J/\psi, \eta_c,
\chi_c, h_c, \cdots; B_s, B_s^*, \cdots$.

\begin{figure}
\centering
%\includegraphics[width=0.280\textwidth]{pic1.eps}%
%\hspace{25mm}\psfrag{g}{\tiny{$J/\psi,\cdots$}}
%\includegraphics[width=0.560\textwidth]{fig1.eps}\hspace*{\fill}
\hspace{25mm}%\psfrag{g}{\tiny{$J/\psi,\cdots$}}
\includegraphics[width=.66\textwidth]{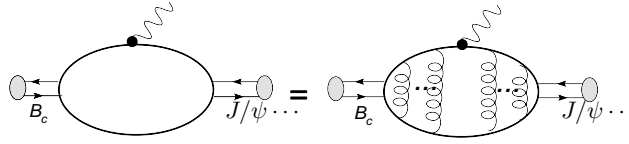}\hspace*{\fill}
\caption{\footnotesize The matrix element of the relevant current
$<P'(B_f)|J_\mu|P(A_i)>$, where the wave line represents a virtual
`particle ($W$-boson)' which brings momentum and quantum number
away; the full dot represents the current which couples to the
virtual $W$-boson; the ladder-like spring-lines in right hand
diagram mean the binding interactions multi-times between the two
components.} \label{fig1} \vspace{-0mm}
\end{figure}

Since the $B_c$ meson is much heavier than the bound state $B_f$
in the final state, so there may be a great momentum recoil, i.e.,
the velocity $v$ of the produced bound state $B_f$ in C,M.S. of
$B_c$ meson may be very great, e.g., $v$ even may reach to $0.7$
(in unit $c$) in the semileptonic decay $B_c\to J/\psi +l+\nu$.
Therefore, the recoil effects should be treated carefully in
estimation of the decays.

To deal with the recoil effects in the decays, there are several
ways, but I think that all of them should `contain'
multi-interactions between the components as indicated in Fig.1
(the right hand diagram). In Refs.\cite{dec,changgl,vary,Gem} the
authors consider the multi-interactions to dictate the recoil
effects with Bether-Salpeter or Dyson-Schwinger equations, while
in Ref.\cite{Russ} the authors consider them with QCD sum rule
with infinite `Coulomb gluon' exchanges. The results obtained by
the different ways are different due to the fact that each of them
has taken certain approximations and has different input
parameters, although all are similar in the inspirits.

\begin{figure}
\centering
%\includegraphics[width=0.380\textwidth]{z1pku.eps}%
%\hspace{25mm}\psfrag{g}{\tiny{$J/\psi,\cdots$}}
%\includegraphics[width=0.560\textwidth]{fig1.eps}\hspace*{\fill}
\hspace{25mm}
\includegraphics[width=.36\textwidth]{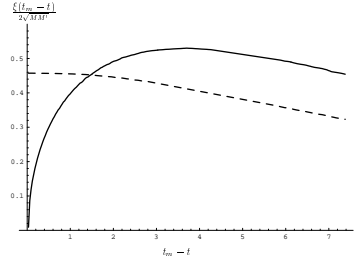}\hspace*{\fill}
\caption{\footnotesize The two `Isgur-Wise'-like functions for the
decays from $S$-wave state ($B_c$ meson) to a $P$-wave state (e.g.
$\eta_c, J/\psi, \cdots$).} \label{fig1} \vspace{-0mm}
\end{figure}

\vspace{-4mm}
\begin{table}[h]
\tbl{Typical semileptonic and nonleptonic decay rates for $B_c$
meson in units of $10^{-15}$ GeV.} {\begin{tabular} {c c c c c ||
c c c c}\hline\hline
%%%%%%    1 row
& $\mathcal{A}$ & $\mathcal{B}$ & $\mathcal{C}$ & $\mathcal{D}$ &
& $\mathcal{A}$ & $\mathcal{C}$ &
$\mathcal{D}$  \\
\hline $B_c\to \eta_c e\nu$         &$14.2$ &$10.7$ & $11.1$
&$11\pm 1$ &  $B_c\to \eta_c \pi$ &$3.29$ &$2.52$ &$\times$\\
\hline $B_c\to \j/\psi e\nu$  & $34.4$ & $28.2$ & $30.2$ &$28\pm
5$ & $B_c\to \j/\psi \pi$  & $3.14$ & $1.94$ &$\times$
\\\hline $B_c\to \chi_{c0} e\nu$ &$1.69$ &$2.52$ &$\times$& $\times$&
 $B_c\to \eta_{c} \rho$ &$8.70$ &$5.94$ & $\times$\\
\hline $B_c\to \chi_{c1} e\nu$ &$2.21$ &$1.40$ &$\times$ &$\times$
&$B_c\to J/\psi \rho$ &$9.45$ &$5.52$ & $\times$ \\ \hline $B_c\to
\chi_{c2} e\nu$ &$2.73$ &$2.92$ &$\times$ & $\times$ &$B_c\to B_s
\pi$ &$73.3$ &$25.1$ & $25.1$ \\
\hline $B_c\to h_c e\nu$ &$2.51$ &$4.42$ &$\times$ &$\times$ &
$B_c\to B_s^* \pi$
&$64.7$ &$19.8$ & $9.84$\\
\hline $B_c\to B_s e\nu$ &$26.6$ &$\times$ & $14.3$ &$58.0$  &
$B_c\to B_s
\rho$ &$56.1$ &$62.2$ & $10.6$\\
\hline $B_c\to B_s^* e\nu$ &$44.0$ &$\times$ & $50.4$ &$72.0$ &
$B_c\to B_s^* \rho$ &$188.$ &$271.$ & $31.8$
\\\hline\hline
\end{tabular}
} %\label{t3}
\end{table}\vspace{-4mm}
\begin{minipage}{119mm}
\noindent {\footnotesize In Table.2 the value of the CKM matrix
element is taken as $|V_{cb}|=0.04$ and the coefficient in the
effective Lagrangian $a_1=1.26$, the results in $\mathcal{A}$
column is taken from Ref.\cite{dec,changgl}, those in column
$\mathcal{B}$ from Ref.\cite{Gem}, those in column $\mathcal{C}$
from Ref.\cite{vary}, those in column $\mathcal{D}$ from
Ref.\cite{Russ}.}
\end{minipage}
\vspace{2mm}

It is interesting to point out here that due to the effects of the
great momentum recoil, for the decays from $S$-wave ($B_c$ meson)
to a $P$-wave state (e.g. $\eta_c, J/\psi, \cdots$) there are two
`Isgur-Wise'-like functions (see Fig.2), while for the decays from
$S$-wave ($B_c$ meson) to an $S$-wave state (e.g. $\eta_c, J/\psi,
\cdots$) there is only one `Isgur-Wise'-like function. The
so-called `Isgur-Wise'-like function(s) means that all of the form
factors for the decays may always be depicted by the function(s)
with proper kinematic factors as in the cases of HQET. The dotted
curve in Fig.2 represents the `common' one which is similar to
that for the decays from $S$-wave to the $S$-wave state and
decreases slowly with the momentum recoil $t_m-T$. Whereas the
solid curve is a `fresh' `Isgur-Wise'-like function, which is zero
at $t_m-t=0$ (null recoil) and increases with the momentum recoil
$t_m-t$ rapidly then goes down slowly. If the recoil effects had
been ignored, then the fresh `Isgur-Wise'-like function would
always is zero. Typical results of the semileptonic decays are put
in Table 2 and we can see that the decay rates of the channels
with $B_s$ or $B_s^*$ in final state are great.

The theoretical estimates on the nonleptonic decays of $B_c$ meson
with naive factorization can be found in many references such as
Refs.\cite{dec,changgl,cchq,vary,Gem} etc. It is too long to
present all of the results here, so alternatively, I only choose
some of them to put in Table 2. One can see the general feature of
the nonleptonic decays from Table 2 that the decay rates of the
channels with $B_s$ or $B_s^*$ in final state are also
comparatively great. \vspace{-4mm}

\section{Production (at Tevatron vs at LHC)}

The reason why the first experimental observation of the `usual'
meson $B_c$ happened so late in 1998 is because the difficulty of
its production i.e. smallness of its production cross-section.
Only at high energy hadronic colliders can the cross-section be
sizable enough for observation, so its first observation was
successful at Tevatron by CDF collaboration. According to
theoretical estimates and the design luminosity of various kinds
of colliders, accurately experimental studies of the meson under
high statistics are accessible only at Tevatron and LHC.

Since the meson $B_c$ carries two heavy flavors explicitly, so to
produce it in the most favorable manner is to produce two pairs of
heavy quarks: $c, \bar{c}, b, \bar{b}$ first, and then the two
heavy quarks $c, \bar{b}$ of them combine into the meson by
certain possibility. Two pairs of the quarks $c, \bar{c}, b,
\bar{b}$ are so heavy that their production always in the
perturbative region of QCD, while the `possibility' to combine the
two heavy quarks $c, \bar{b}$ into the meson, being
nonperturbative nature, relates to a relevant matrix element in
NRQCD framework (for color-singlet production it can be related to
the wave function of $B_c$ at origin in potential model framework)
directly\cite{changq}. Therefore the production of $B_c$ meson may
be always estimated by perturbative QCD (pQCD) with proper
factorization
formulation\cite{prod,changq,prod0,prod1,prod2,prod5}.

\vspace{-4mm}
\begin{table}[h]
\tbl{Total cross-sections (LO QCD estimate and in unit of nb) for
the hadronic production of $B_c[1^{1}S_{0}]$ and
$B_c^*[1^{3}S_{1}]$ at TEVATRON and at LHC.}
{\begin{tabular}{c||c|c|c||c|c|c} \hline\hline
 & CTEQ6L & GRV98L & MRST2001L&
 CTEQ6L & GRV98L & MRST2001L \\
\hline - &\multicolumn{3}{|c||}{$Q^2=\hat{s}/4$} &
\multicolumn{3}{|c}{$Q^2=p_{T}^2+m_{B_c}^2$}\\
\hline\hline - & \multicolumn{6}{|c }{TEVATRON} \\\hline
$\sigma_{B_c(1^{1}S_{0})}$& 3.79 &
3.27 & 3.40 & 5.50  & 4.54 & 4.86\\
\hline $\sigma_{B^*_c(1^{3}S_{1})}$ & 9.07 &
7.88 & 8.16 & 13.4 & 11.1 & 11.9\\
\hline \hline - & \multicolumn{6}{|c}{LHC}\\\hline
$\sigma_{B_c(1^{1}S_{0})}$ & 53.1 &
53.9 & 47.5 & 71.1 & 70.0 & 61.4 \\
\hline $\sigma_{B^*_c(1^{3}S_{1})}$ & 130. &
131.& 116. & 177.& 172. & 153.\\
\hline \hline
\end{tabular}}
\end{table}
\vspace{-6mm}
\begin{center}
\begin{minipage}{124mm}
{\footnotesize In Table. 3, the characteristic energy scale $Q$ is
taken as $Q^2=\hat{s}/4$ and $\sqrt{\hat{s}}$ is the C.M. energy
of the active subprocess or $Q^2=p_{T}^2+m_{B_c}^2$ and $p_T$ is
the transverse momentum of $B_c$ meson; the values are taken from
Ref.\cite{prod2}.}
\end{minipage}
\end{center}

There are two theoretical approaches of pQCD to the estimate of
its production: the so-called fragmentation approach\cite{prod0}
and the so-called complete calculation approach of the
lowest-order perturbative QCD\cite{changq,prod1,prod2,prod5}.
Since the second one may keep more useful information of the
production for experiments, it is highlighted in literature. There
are several mechanisms for the hadronic production of $B_c$ meson,
in most $P_T$ (transverse momentum of the produced $B_c$ meson)
region, the so-called `gluon-gluon fusion' mechanism is
domimant\cite{prod1,prod2}. Since only LO estimates of $B_c$
production are available up-to now, so several important
theoretical uncertainties for LO estimate are investigated in
Ref.\cite{prod2}. Recently the $P$-wave excited $B_c$ production
not only via its color-singlet component but also via its
color-octet components is estimated, and it is pointed out that
the $P$-wave production can be sizable\cite{prod5}. The general
feature of the production is the cross sections increase slowly
with the center mass energy of the collision $\sqrt S$. The total
production cross-sections\cite{prod2} at Tevatron and LHC are put
in Table. 3 and the precise $p_T$ distributions for the production
can be found in Refs.\cite{prod2,prod5}. From Table. 3, one still
can see that the cross-sections at LHC are greater than those at
Tevatron by one order of magnitude, therefore, the studies at LHC
can have much higher statistics than that at Tevatron.

It is remarkable that to meet various experimental needs of event
simulation for feasibility studies of the meson $B_c$ at Tevatron
and LHC, a computer program (the event generator for $B_c$) named
BCVEGPY that is in compliment to the PYTHIA environment has been
completed\cite{bcvegpy1}. It is powerful enough for most purposes,
i.e., with it one can enhance the event generating efficiency
greatly in contrast to PYTHIA itself. With the latest version
BCVEGPY2.0, not only the ground state of $B_c$ meson can be
generated, but can also its $P$-wave excited states.\vspace{-4mm}

\section{Outlook}

In prospects of the cross sections of $B_c$ production, available
detectors, collision luminosity at Tevatron (Run-II) and at LHC,
accurately and thoroughly experimental studies of it is accessible
very soon. $B_c$ physics is compelling.  With very high statistics
we will have various tests of the theoretical estimates
(predictions) etc. and might have better measurements on CKM
matrix elements. The self-tagged $B_s$ mesons via $B_c$ decays as
a tagged $B_s$ source might be achieved. The future copious data
require more accurate theoretical predictions in the environments
at Tevatron and at LHC respectively since now on etc. The studies
of $B_c$ meson will jump to a new stage not only for itself but
also to implement the studies of the heavy quarkonia etc.

\section*{Acknowledgments}

I would like to thank the organizers of the conference very much
for their kind invitation. This work is supported by Natural
Science Foundation of China (NSFC).

\end{document}